\documentclass{epl}
\usepackage{amsmath}
\usepackage{amssymb}
\usepackage{graphicx}

\title{A Yule-Simon process with memory}
\author{C.~Cattuto\inst{1,2} \and V.~Loreto\inst{2} \and V.~D.~P. Servedio\inst{3,1}}

\institute{                    
  \inst{1} Museo Storico della Fisica e Centro
Studi e Ricerche Enrico Fermi , Compendio Viminale, 00184 Rome, Italy\\
  \inst{2} Dipartimento di Fisica, Universit\`a ``La Sapienza'',
P.le A. Moro 2, 00185 Roma, Italy\\
  \inst{3} Dipartimento di Informatica e Sistemistica, Universit\`a
``La Sapienza'', Via Salaria 113, 00198 Roma, Italy
}
\pacs{05.10.-a}{Computational methods in statistical physics and nonlinear dynamics}
\pacs{05.40.-a}{Fluctuation phenomena, random processes, noise, and Brownian motion}
\pacs{89.20.Ff}{Computer science and technology}

\newcommand{\bp}{\bar{p}}

\begin{document}

\maketitle

\begin{abstract}
The Yule-Simon model has been used as a tool to describe the growth of
diverse systems, acquiring a paradigmatic character in many fields of research.
Here we study a modified Yule-Simon model that takes into account the
full history of the system by means of an hyperbolic memory kernel.
We show how the memory kernel changes the properties of preferential attachment
and provide an approximate analytical solution for the frequency
distribution density as well as for the frequency-rank distribution.
\end{abstract}


In 1925 Yule~\cite{yule_1925} proposed a model to explain experimental
data on the abundances of biological genera~\cite{willis_1922}.  Thirty
years later, Simon introduced an elegant copy and growth
model~\cite{simon_1955}, in spirit equivalent to Yule's model, to
explain the observed power-law distribution of word frequencies in
texts~\cite{estoup_1916, condon_1928, zipf_1935}.  In Simon's growth
model, new words are added to a text (more generally a stream) with
constant probability $p$ at each time step, whereas with complementary
probability $\bar{p}=1-p$ an already occurred word is chosen uniformly
from within the already formed text (stream).  This model yields a
power-law distribution density for word frequencies $P(k)\sim k^{-\beta}$ with
$\beta=1+1/\bp$ . The same mechanism is at play in the preferential
attachment (PA) model for growing networks proposed, in their
pioneering article, by Barab\'asi and Albert~\cite{BA}.  In that
case, a network is constructed by progressively adding new nodes and
linking them to existing nodes with a probability
proportional to their current connectivity.  Yule-Simon processes and
PA schemes are closely related to each other
and a mapping between them has been provided by
Bornholdt and Ebel~\cite{bornholdt}.

In the original Yule-Simon process, the metaphor of text construction
is somehow misleading because in that process there is no notion of
temporal ordering. All existing words are equivalent and in many
respects everything goes as in a Polya urn model~\cite{johnson}.
However, the notion of temporal ordering may play an important role in
determining the dynamics of many real systems. In this perspective it
is interesting to investigate models where temporal ordering is
explicitly taken into account.  A first attempt in this direction has
been provided by Dorogovtsev and Mendes (DM)~\cite{mendes}, who
studied a generalization of the Barab\'asi-Albert model by introducing
a notion of aging for nodes. Each node carries a temporal marker
recording its time of arrival into the network, and its probability to
be linked to newly added nodes is proportional to its current
connectivity weighted by a power-law of its age.  Another recent
example has been proposed in~\cite{clp_2006} in relation with the very
new phenomenon of collaborative tagging~\cite{huberman}: new web sites
appeared where users independently associate descriptive keywords --
called tags -- with disparate resources ranging from web pages to
photographs.  A sort of tag dynamics develops, eventually yielding a
fat-tailed distribution of tag frequencies.  In order to explain such
phenomenology, a generalization of the Yule-Simon process has been
introduced~\cite{clp_2006}, which explicitely takes into account the
time ordering of tags. Specifically, an hyperbolic memory kernel has
been introduced to weight the probability of copying an existing tag,
affording a remarkable agreement with experimental data.

In this Letter we show that the memory kernel
induces a non-trivial change of the properties of PA with respect to
the original Yule-Simon process as well as to the DM model with aging.
Moreover, we analytically investigate the generalization of the Yule-Simon
model and provide an approximate solution for the frequency
distribution density as well as for the frequency-rank distribution.


\begin{figure}
\onefigure[width=0.75\columnwidth]{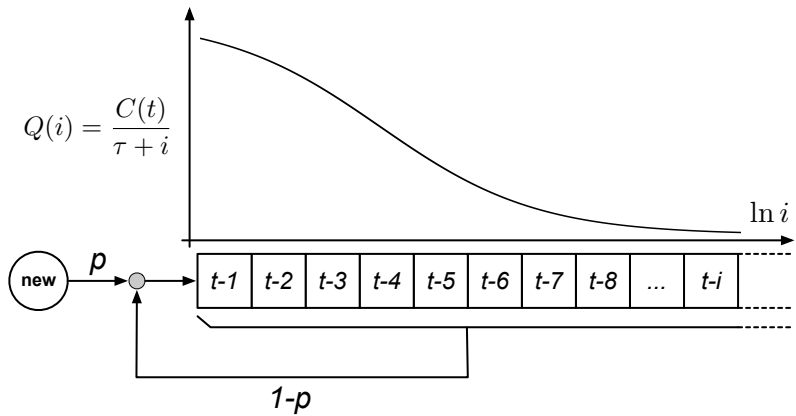}
\caption{Yule-Simon process with a fat-tailed memory kernel.}
\label{fig:model}
\end{figure}

The model we investigate is defined as follows. We start with $n_0$
words. At every time step $t$ a new word may be
invented with probability $p$ and appended to the text, while with
probability $\bar{p}=1-p$ one word is copied from the text, going back
in time by $i$ steps with a probability that decays with $i$ as \(
Q(i) = \frac{C(t)}{\tau + i} \, \), as shown in Fig.~\ref{fig:model}.
$C(t)$ is a logarithmic time-dependent normalization factor
and $\tau$ is a characteristic time-scale over which recently added words
have comparable probabilities.

\begin{figure}
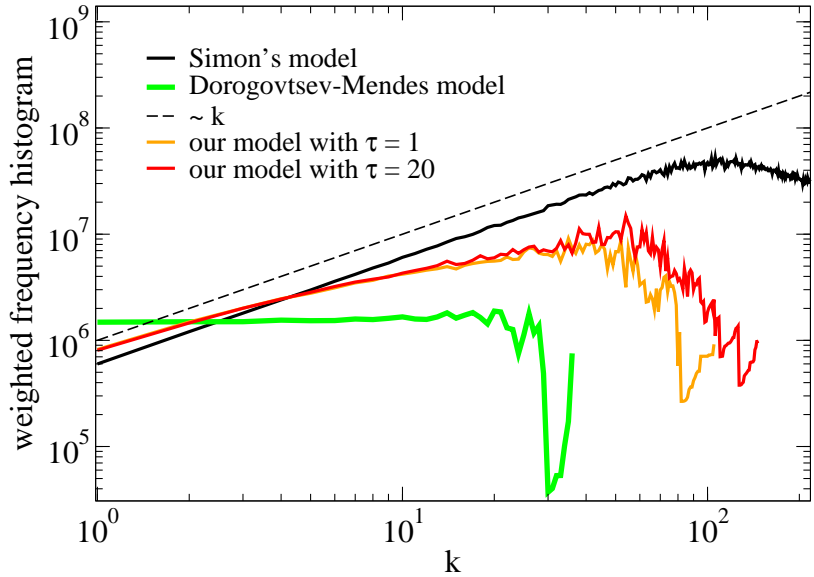

\onefigure[width=0.75\columnwidth]{figure2}
\caption{Deviations from the preferential attachment rule (Simon's model),
in the case of our model and DM model.
For all curves, $p=0.4$ and $10^6$ steps were simulated.
Finite size effects are responsible for the drop at high frequencies,
as extensively discussed in Ref.~\cite{newmanPAC}.}
\label{fig:PAC}
\end{figure}

The first important observation concerns the deviations of our model
from the pure PA rule of the original Yule-Simon model.  An elegant
and efficient way to check for deviations from PA was suggested by
Newman \cite{newmanPAC}.  In Simon's model, the probability of choosing
an existing word, which already occurred $k$ times at time $t$, is
$\bar{p}\,k\,\pi(k,t)$, where $\pi(k,t)$ is the fraction of words with
frequency $k$ at time $t$.  In order to ascertain whether a PA
mechanism might be at work, we construct the histogram of the
frequencies of words that have been copied, weighting the contribution
of each word according to the factor $1/\pi(k,t)$. If this histogram
displays a direct proportionality to the frequency $k$, then one might
be observing a PA-driven growth.  For our model, the numerical results
in Fig.~\ref{fig:PAC} show that the chosen form of the memory kernel
leads to a sub-linear attaching probability. The same kind of
sub-linearity has been observed in the growth dynamics of the
wikipedia network~\cite{capocci}. Conversely, the DM model with
hyperbolic kernel (a limiting case for the analysis of
Ref.~\cite{mendes}) displays no clear dependence on $k$.

In order to get a deeper insight into the phenomenology of the model
we present an analytical study aimed at computing the approximate
functional form of the probability distribution of word frequencies as
well as the corresponding frequency-rank distribution. In the
following we shall write the normalization factor as $C$, with no
explicit mention of its time dependence. We also define \(\alpha(t)
\equiv \bp \, C(t)\), and we will similarly refer to it as $\alpha$.
We assume that word $X$ occurred at time $t$ for the first time, and
we ask what is the probability $P(\Delta t)$ that the next occurrence
of $X$ happens at time $t+\Delta t$, with $\Delta t \geq 1$.

If $\Delta t = 1$, $P(\Delta t)$ is the probability of replicating the
previous word, i.e. the product between the probability $\bar{p}$ of
copying an old word, and the probability of choosing the immediately
preceding word ($i = 1$) computed according to the chosen memory
kernel, $Q(1) = C / (\tau +1)$.  This gives
\begin{equation}
P(1) = \frac{\bar{p}\,C}{\tau + 1} = \frac{\alpha}{\tau + 1}  \, .
\label{eq:return1}
\end{equation}

For $\Delta t > 1$, $P(\Delta t)$ can be computed as the product of the
probabilities of \textit{not} choosing word $X$ for $\Delta t - 1$
consecutive steps, multiplied by the probability of choosing word $X$
at step $\Delta t$.  In order not to choose word $X$ at the first
step, one has to either append a new word (probability $p$) or copy an
existing word (probability $\bar{p}$) which is not $X$ (probability $1 - C
/ (\tau + 1)$). 

Finally, under the approximation that $C$ is constant from step to
step, i.e. $\Delta t \ll t$, we can write the return probability as
the product

\begin{equation}
	P(\Delta t) \simeq \left[ \frac{\bar{p}\,C}{\tau + \Delta t}
	 \right] \cdot \prod_{i=1}^{\Delta t -1} \left[ p + \bar{p}
	 \left(1 - \frac{C}{\tau + i} \right) \right] \, .
\end{equation}

\noindent
Taking the logarithm of $P(\Delta t)$, we can write the above product
as the sum
\begin{equation}
\label{eq:return2}
	\ln P(\Delta t) =
	 -\alpha \sum_{i=1}^{\Delta t -1}
	 \frac{1}{\tau + i} + \ln \frac{\alpha}{\tau + \Delta t} \, ,
\end{equation} 
where we used the fact that $\alpha\ll 1$ for $t \gg 1$.

By using the approximate expression
\( \ln P(\Delta t) = \int_1^{\Delta t} (\ln P(\Delta t' +1) - \ln P(\Delta t')) \, d\Delta t' \)
we obtain 
\begin{equation}
  P(\Delta t) \simeq \alpha (1+\tau)^\alpha (\tau+\Delta t)^{-\alpha -1}
  \label{eq:Pdt}
\end{equation}
derived under the assumption that $t \gg \Delta t \gg 1$.  The
estimated value of $P(\Delta t)$ depends on time through $\alpha$, so
that the probability distribution of intervals $\Delta t$ ,
which turns out to be correctly normalized, is non-stationary.\\ 
We now focus,
for simplicity, on the case $\tau = 0$.  At any given time $t$, the
characteristic return time $\langle\Delta t\rangle$ can be computed by
using Eq.~\ref{eq:Pdt}:
\begin{equation}
\label{eq:return_avg}
\langle\Delta t\rangle = {\sum_{\Delta t =1}^{t} \, \ P(\Delta t)
\, \Delta t}  \simeq
\frac{\alpha}{1 -\alpha} \, t^{1-\alpha} \,\, .
\end{equation}

In a continuum description the frequency $k_i$ of a given word $i$, 
will change according to the rate equation
\begin{equation}
\frac{dk_i}{dt} = \bp \, \Pi_i \, ,
\label{eq:13}
\end{equation}
where $\Pi_i$ is the probability of picking up a previous occurrence
of word $i$.
With our choice of the memory kernel, the exact value of $\Pi_i$ is
given by the sum
\begin{equation}
\Pi_i = C \, \sum_{j=1}^{j=k_i} \, \frac{1}{t - t^{(i)}_j} \, ,
\label{eq:rate}
\end{equation}
where $t^{(i)}_j$ ($j = 1, 2, \ldots , k_i$) are the times of
occurrence of word $i$.

We adopt a mean-field approach and assume that the above sum can be
written as the frequency $k_i$ times the average value of
the term $(t-t^{(i)}_j)^{-1}$ over the occurrence times $t^{(i)}_j$.

\begin{figure}
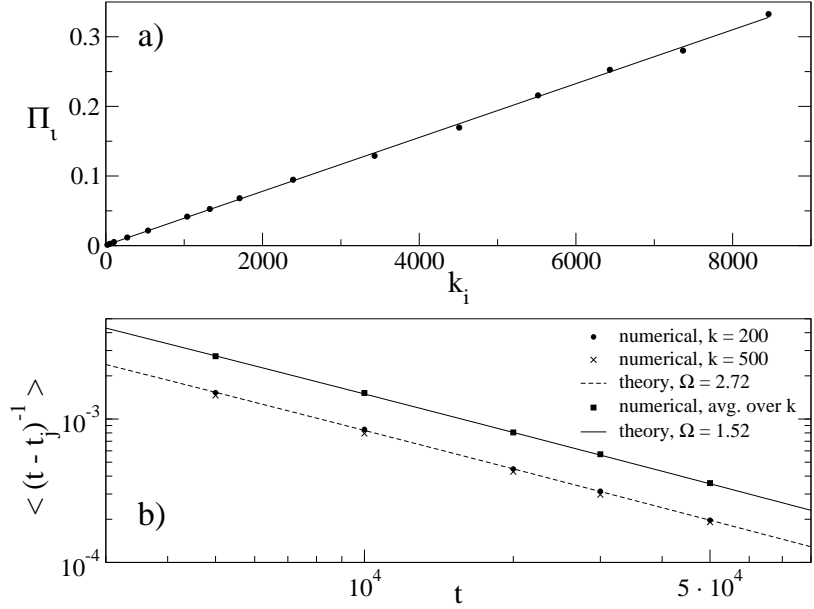

\onefigure[width=0.75\columnwidth]{figure3}
\caption{{\bf a)} Rate $\Pi_i$ of Eq.~\protect{\ref{eq:rate}}, for a
given word $i$ having frequency $k_i$ at time $t$ ($p=0.05$, $n_0=10$,
$t=30000$) {\bf b)} Memory kernel of
Eq.~\protect{\ref{eq:kernel_avg}} averaged over the times of
occurrence $t_j$ and over about $2000$ realizations of the process
($p=0.05$, $n_0=10$,
$t=5\cdot10^3,10^4,2\cdot10^4,3\cdot10^4,5\cdot10^4$).  Values are
shown for a word of given frequency $k=200$ (dots), a word of
frequency $k=500$ (crosses) and for the average over all frequencies
(squares). Numerical error bars are within the size of data
markers. The two curves are obtained by fitting $\Omega$ in
Eq.~\protect{\ref{eq:kernel_avg}} against numerical data.
The fitted continuous line sets the value of $\Omega$ used
from Eq.~\protect{\ref{eq:rate2}} onwards.}
\label{fig:kernel_panels}
\end{figure}

As shown in Fig.~\ref{fig:kernel_panels}a, this is supported by numerical
evidence, so that we can write (dropping the word index $(i)$ from here onward):
\begin{equation}
\Pi_i = C \, \sum_{j=1}^{j=k_i} \, \frac{1}{t - t_j} \simeq C \, k_i
 \, \left< \frac{1}{t - t_j} \right>_{j} \, ,
\label{eq:15}
\end{equation}
where $\langle \, \rangle_j$ denotes the average over the $k_i$
occurrences of word $i$.  Furthermore, we assume that the average is
dominated by the contribution of the most recent occurrence of word
$i$, at time $t_{k_i}$: \( \langle (t - t_j)^{-1} \rangle_{j} \simeq
(t - t_{k_i})^{-1} \).  We replace $t - t_{k_i}$ with the typical
return interval for word $i$, and use Eq.~\ref{eq:return_avg} to
estimate the latter, obtaining:
\begin{equation}
\left< \frac{1}{t - t_j} \right>_{j} \simeq \frac{1}{t - t_{k_i}}
\simeq \frac{1}{\langle \Delta t \rangle} = \frac{1-\alpha}{\alpha}
\cdot \frac{1}{t^{1-\alpha}} \, ,
\label{eq:kern_estimate}
\end{equation}
which has a (sub-linear, since $\alpha \gtrsim 0$) power-law
dependence on $t$ and a slower (logarithmic) time-dependence through
$\alpha$.  Fig.~\ref{fig:kernel_panels}b shows that the above
expression captures the correct temporal dependence of the average
$\langle (t - t_j)^{-1} \rangle$ for a given frequency $k_i$, provided
that a constant factor $\Omega$ is introduced, as follows:
\begin{equation}
\left< \frac{1}{t - t_j} \right>_{j} \simeq \frac{1}{\Omega} \cdot
\frac{1-\alpha}{\alpha} \cdot
\frac{1}{t^{1-\alpha}} \, .
\label{eq:kernel_avg}
\end{equation}
The need for a corrective factor $\Omega$ is a consequence of our
simplifying assumptions, namely our mean-field approximation, the fact
that we ignored all occurrences of word $i$ but the very last, and the
approximations underlying our estimate of the return time $\Delta t$.
Moreover, as shown in Fig.~\ref{fig:kernel_panels}b, $\Omega$ depends
on the frequency $k_i$ of the selected word $i$.
In order to keep only the linear dependence
of the kernel on $k_i$ we approximate $\Omega$ with its average value
over $k$, numerically estimated as $\Omega \simeq 1.52$ (see
Fig.~\ref{fig:kernel_panels}b). While this is certainly a rather crude
approximation, it appears to work remarkably well, as we will show
in the following (Figs.~\ref{fig:pdk} and \ref{fig:pdr}).\\
We introduce Eq.~\ref{eq:kernel_avg} and Eq.~\ref{eq:15}, into
the rate Eq.~\ref{eq:13}, obtaining:
\begin{equation}
\frac{dk_i}{dt} \simeq \alpha \, k_i \, \left< \frac{1}{t - t_j}
\right>_{j} = \frac{k_i}{\Omega} \cdot (1 - \alpha) \cdot t^{\alpha-1}
\, .
\label{eq:rate2}
\end{equation}
We integrate Eq.~\ref{eq:rate2}, again neglecting the slow time-dependence
of $\alpha$, from time $t_i$, when word $i$ appeared for the
first time (with frequency $1$) up to the final time $t$, when word $i$
has reached frequency $k_i$,
\begin{equation}
\int_{1}^{k_i} \frac{dk^\prime_i}{k^\prime_i} = \frac{1-\alpha}{\Omega
} \cdot \int_{t_i}^{t} dt^\prime \,
{t^\prime}^{\alpha-1} \, .
\end{equation}
Performing the integration we get the stretched exponential dependence
\begin{equation}
k_i = \exp\left[ \frac{1-\alpha}{\Omega \, \alpha } \, t^\alpha \right] \cdot \exp\left[ -
 \frac{1-\alpha}{\Omega \, \alpha } \, t_i^\alpha
 \right] = A \, e^{- K t_i^\alpha} \, ,
\label{eq:k_vs_t}
\end{equation}
where
\(K \equiv \frac{1-\alpha}{\Omega \, \alpha } \) and 
\(A \equiv e^{ K t^\alpha }\).

The probability distribution density for word frequencies $P(k)$ can
now be computed as~\cite{barabasi_review, dorog_book}:
\begin{equation}
P(k) = \frac{p}{(n_0 + pt) \, (K
\alpha) \, k} \, \left[ \frac{\ln(A / k)}{K} \right]^{\frac{1}{\alpha}
- 1} \, ,
\label{eq:pdk}
\end{equation}
and is in very good agreement with numerical evidence,
as shown in Fig.~\ref{fig:pdk} (upper curves), where it is worth
noticing that the value of $\Omega$ is univocally set by our numerics.
The corresponding frequency-rank distribution is:
\begin{equation}
P(R) \simeq \frac{A}{n_0 + t} \, \exp{\left[-K \left( \frac{R}{p}
\right)^\alpha \right]} \, .
\label{eq:pdr}
\end{equation}
Fig.~\ref{fig:pdr} shows that the above equation is in fair
agreement with numerical evidence. 
\begin{figure}
\twofigures[height=4.75cm]{figure4}{figure5}
\caption{Frequency probability distribution density $P(k)$ of word occurrence.
Numerical data (dots, averaged over $50$ realizations and binned)
are in very good agreement with Eq.~\protect{\ref{eq:pdk}} (solid line)
($p=0.05$, $n_0=10$, $t=30000$, $\Omega = 1.52$).
The dashed line is provided as a guide for the eye.}
\label{fig:pdk}
\caption{Frequency-rank distribution $P(R)$.
Upper curves: Numerical data (dots, average over $50$ realizations)
are compared against the prediction of Eq.~\protect{\ref{eq:pdr}} (solid line)
($p=0.05$, $n_0=10$, $t=30000$, $\Omega = 1.52$).
Here the value of $\Omega$ is univocally set by our numerics,
as explained in Fig.~\protect{\ref{fig:kernel_panels}}b.
Lower curves (shifted one decade downwards): a single realization of our process (squares)
is fitted with respect to $\Omega$ against Eq.\protect{\ref{eq:pdr}} (dashed line),
yielding $\Omega = 1.46$.}
\label{fig:pdr}
\end{figure}


In this Letter we have shown how the introduction of a memory kernel
drastically changes the properties of PA with respect of the original
Yule-Simon process as well as the Dorogovtsev-Mendes model with
aging~\cite{mendes}.

In order to assess the role of the memory kernel we have presented a
continuum approach to a modified Yule-Simon model.  The presence of a
long-term memory kernel makes the rigorous treatment non trivial. Our
approach makes use of some assumptions (sometimes rough, but
numerically verified) concerning especially the functional form of the
averaged memory kernel, both as a function of time and of word
frequency.  We require a single phenomenological parameter ($\Omega$),
for which we presently have no theoretical estimates.  Nevertheless
our approach affords an excellent agreement between analytical and
numerical results for the probability distribution density $P(k)$.
The frequency-rank distribution $P(R)$ appears to be much more
sensitive to the approximations we made, but the agreement between
numerics and theory is nevertheless reasonable.  This is somehow the
signature that our theoretical treatment is capturing some of the
important statistical features of the model.

We wish to remark that the frequency probability density $P(k)$
displayed by the model (Fig.~\ref{fig:pdk}) could be easily confused
with a power-law behavior with exponent $-2$, as in the original
Yule-Simon model with $p \ll 1$, and a simple PA mechanism could be
inferred. Instead, as shown for the case at hand, more refined
indicators (e.g.\ that of Fig.~\ref{fig:PAC}) can tell apart different
underlying mechanisms of growth.  This should be read as a general
warning against reading an apparent power-law behavior for the $P(k)$
as the signature of a PA mechanism at play.

The approach described here could be extended to the more complex case
of $\tau \neq 0$.  In this respect, several problems remain open: does
$\tau$ induce a relevant time scale?  Is it asymptotically relevant or
does it only affect the dynamics on short time-scales?  Does the limit
$\tau \gg t$ fall in the same universality class of the Yule-Simon
model without memory?

\acknowledgments
This research has been partly supported by the
TAGora and DELIS projects funded by the Future and Emerging
Technologies program (IST-FET) of the European Commission under
the contracts IST-34721 and IST-1907.
The information provided is the
sole responsibility of the authors and does not reflect the
Commission's opinion. The Commission is not responsible for any use
that may be made of data appearing in this publication.


\begin{thebibliography}{0}

\bibitem{yule_1925}
  \Name{Yule G.~U.} 
  \REVIEW{Philos.\ Trans.\ R.\ Soc.\ London B}{213}{21-87}{1925}

\bibitem{willis_1922}
\Name{Willis J.~C.}
\Book{Age and area: a study in geographical distribution and origin of species}
\Publ{Univ.\ Press, Cambridge}
\Year{1922}

\bibitem{simon_1955} 
\Name{Simon H.~A.}
\REVIEW{Biometrika}{42}{425}{1955}
	
\bibitem{estoup_1916}
\Name{Estoup J.~B.}
\Book{Gammes st\'{e}nographique}
\Publ{Institut St\'{e}nographique de France, Paris}
\Year{1916}

\bibitem{condon_1928}
\Name{Condon E.~U.}
\REVIEW{Science}{67}{300}{1928}

\bibitem{zipf_1935}
\Name{Zipf G.}
\Book{The Psycho-Biology of Language}
\Publ{Houghton Mifflin, Boston, MA}
\Year{1935}

\bibitem{BA} 
\Name{Barab\'asi A.-L. \and Albert R.} 
\REVIEW{Science}{286}{509}{1999}

\bibitem{bornholdt} 
\Name{Bornholdt S. \and Ebel H.} 
\REVIEW{Phys. Rev. E}{64}{035104}{2001}

\bibitem{johnson} 
\Name{Johnson N.~L. \and Kots S.} 
\Book{Urn models and their application}
\Publ{John Wiley, New York}
\Year{1977}

\bibitem{mendes} 
\Name{Dorogovtsev S.~N. \and Mendes J.~F.~F.} 
\REVIEW{Phys. Rev. E}{62}{1842}{2000}

\bibitem{clp_2006} 
\Name{Cattuto C., Loreto V. \and Pietronero L.}
\Review{cs/0605015 preprint}
\Year{2006}

\bibitem{huberman}
\Name{Golder S. \and Huberman B.~A}
\REVIEW{J. of Information Science}{32}{198}{2006}

\bibitem{newmanPAC} 
\Name{Newman M.~E.~J.}
\REVIEW{Phys. Rev. E}{64}{025102R}{2001}

\bibitem{capocci} 
\Name{Capocci A., Servedio V.~D.~P., Colaiori F., Buriol L.~S.,
Donato D., Leonardi S. \and G.~Caldarelli}
\Review{physics/0602026 preprint}
\Year{2006}

\bibitem{barabasi_review} 
\Name{Albert R. \and Barab\'asi A.-L.}
\REVIEW{Rev.\ Mod.\ Phys.}{74}{47}{2002}

\bibitem{dorog_book} 
\Name{Dorogotsev S.~N. \and Mendes J.~F.~F.}
\Book{Evolution of Networks: From Biological Nets to the Internet and WWW}
\Publ{Oxford University Press, Oxford} 
\Year{2003}

\end{thebibliography}
\end{document}